\documentclass[fleqn,12pt]{article}
\usepackage[dvips]{graphicx}
\usepackage{cite}
\usepackage{times}
\renewcommand{\refname}{\normalsize References}

\makeatletter
\renewenvironment{thebibliography}[1]
{\section*{\refname\@mkboth{\refname}{\refname}}\vspace{4pt}%
   \list{\@biblabel{\@arabic\c@enumiv}}%
        {\settowidth\labelwidth{\@biblabel{#1}}%
         \leftmargin\labelwidth
         \advance\leftmargin\labelsep
	 \setlength\itemsep{0pt}%
	 \setlength\baselineskip{14pt}%
         \@openbib@code
         \usecounter{enumiv}%
         \let\p@enumiv\@empty
         \renewcommand\theenumiv{\@arabic\c@enumiv}}%
   \sloppy
   \clubpenalty4000
   \@clubpenalty\clubpenalty
   \widowpenalty4000%
   \sfcode`\.\@m}
  {\def\@noitemerr
    {\@latex@warning{Empty `thebibliography' environment}}%
   \endlist}
 
\makeatletter
\long\def\abstract#1{\def\@abstract{#1}}%
\def\@abstract{}
\long\def\affiliation#1{\def\@affiliation{#1}}%
\def\@affiliation{}
\long\def\email#1{\def\@email{#1}}%
\def\@email{}
\def\@maketitle{%
\begin{center}
\vspace*{1pt}
{\large\bfseries\@title}
\\[12pt]
\@author
\\[12pt]
{\textit{\@affiliation}}
\\
\@email
\end{center}
\leftmargini15mm
\begin{quotation}\noindent {\bfseries\abstractname}\\
\@abstract\end{quotation}%
 \vskip14pt}%
\makeatother

\makeatletter
\renewcommand{\section}{\@startsection
{section}{3}{0mm}{5mm}{0.1pt}{\bfseries \normalsize}}
\makeatother

\makeatletter
\renewcommand{\subsection}{\@startsection
{subsection}{3}{0mm}{5mm}{0.01pt}{\bfseries \normalsize}}
\makeatother

\title{%
Precise measurement of Hyper Fine Structure of positronium using sub-THz light.
}

\author{%
S.Asai$^1$, T.Suehara$^1$, T.Yamazaki$^1$, G.Akimoto$^1$, A.Miyazaki$^1$, T.Namba$^1$, T.Kobayashi$^1$,
H.Saito$^2$, T.Idehara$^{3}$, I.Ogawa$^{3}$, Y.Urishizaki$^{3}$
~and
S.Sabchevski$^4$
}

\affiliation{%
$^1$Dept.\ of Physics, and International Center for the Elementary Particle Physics(ICEPP), University of Tokyo, Japan.\\
$^2$Dept.\ of General systems studies, Graduate School of Arts and Sciences, University of Tokyo, Japan.\\
$^3$Research Center for Development of Far-Infrared region, University of Fukui, Japan.\\
$^4$Bulgarian Academy of Science,Sofia,  Bulgaria.
}

\email{%
Shoji.Asai@cern.ch
}

\abstract{
Positronium is an ideal system for the research 
of the QED, especially for the QED in bound state.
The discrepancy of 3.9$\sigma$ is found recently between 
the measured HFS values and the QED prediction ($O(\alpha^3)$).
It might be due to the contribution of the unknown new physics or
the systematic problems in the previous all measurements.
We propose new method to measure HFS precisely and directly.
A gyrotron, a novel sub-THz light source is used with a high-finesse 
Fabry-P\'erot cavity to obtain enough radiation power at 203 GHz.
The present status of the optimization studies and current design of the
experiment are described.
}

\begin{document}
\maketitle
\thispagestyle{empty}
\pagestyle{empty}

\section{Introduction}

Positronium (Ps), the bound state of 
an electron and a positron,
is a purely leptonic system and the triplet ($1^{3}S_{1}$) state of Ps, orthopositronium(o-Ps),
decays slowly into three photons.
The o-Ps has the energy higher than 
the single state ($1^{1}S_{0}$) of Ps, parapositronium(p-Ps), which decays into two photons promptly.
The difference of the energy level between o-Ps and p-Ps is 
called as Hyper Fine Stricture (HFS) and 
is significantly larger(about 203~GHz) than the hydrogen-atom (1.4~GHz) 
because of the following two reasons:
(1) The magnetic moment is proportional to the inverse of the mass,
thus the large spin-spin interaction is expected for o-Ps.
(2) o-Ps has the quantum number the same as photon, then o-Ps makes the quantum oscillation through virtual photon;
o-Ps $\rightarrow \gamma^{*} \rightarrow$ o-Ps. 
This oscillation frequency of 87~GHz contributes only to the o-Ps, 
and makes the HFS larger. 

The precise measurement of the HFS gives the direct information about
the QED, especially of the bound state QED.
If an unknown light particle (like as axion or millichaged particle) exits,
it contributes to the energy level, and makes discrepancy from the QED prediction.
Since the quantum oscillation has good sensitivity to such a hypothesis 
particle, whose coupling is super-weak, 
the precise measurement of HFS is good tool to search for the new physics beyond
the Standard Model indirectly.

The precise measurements have been performed in 70's and 80's.
These results are consistent with each other and the final accuracy is 3.6 ppm~\cite{HFSnew}.
New method to calculate including the higher order correction on the bound state 
is established in 2000, 
and the 2nd and 3rd corrections have been performed\cite{HFSth}.
The QED prediction is 203.3917(6) GHz and it differs from the measured value of
203.3889(67) GHz. 
The discrepancy of 3.9$\sigma$ is observed and
but there would be possibility of the new physics or
the common systematic errors in the previous measurements.

\section{Old experiments and the systematic errors}
 
In the previous all measurements, the HFS transition was not directly measured,
since the 203GHz is too high frequency to be handled.
The static magnetic field makes Zeeman mixing between $m_z$=0 state of o-Ps 
and p-Ps, the resultant energy level of new $m_z$=0 state 
(referred as $|+\rangle$ in figure)  becomes higher than  $m_z=\pm$1. 
This energy shift is proportional to the HFS energy level. 
Figure~1(a) shows the energy levels as a function of the static magnetic field.
On the other hand, the $m_z=\pm$1 state of o-Ps do not couple to the static magnetic field,
and do not change the energy level.
The energy shift between $|+\rangle$ 
and $m_z=\pm$1 is called as the Zeeman shift.
The Zeeman shift, which is proportional to the HFS, has been measured in the previous 
all experiments. 

\begin{figure}[h]
\begin{center}
\includegraphics[height=60mm]{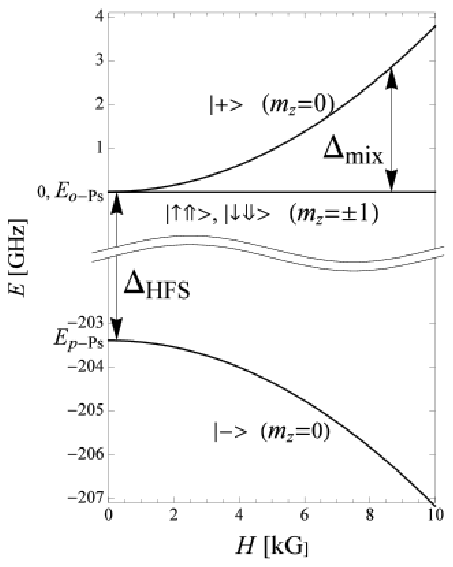}
\includegraphics[height=60mm]{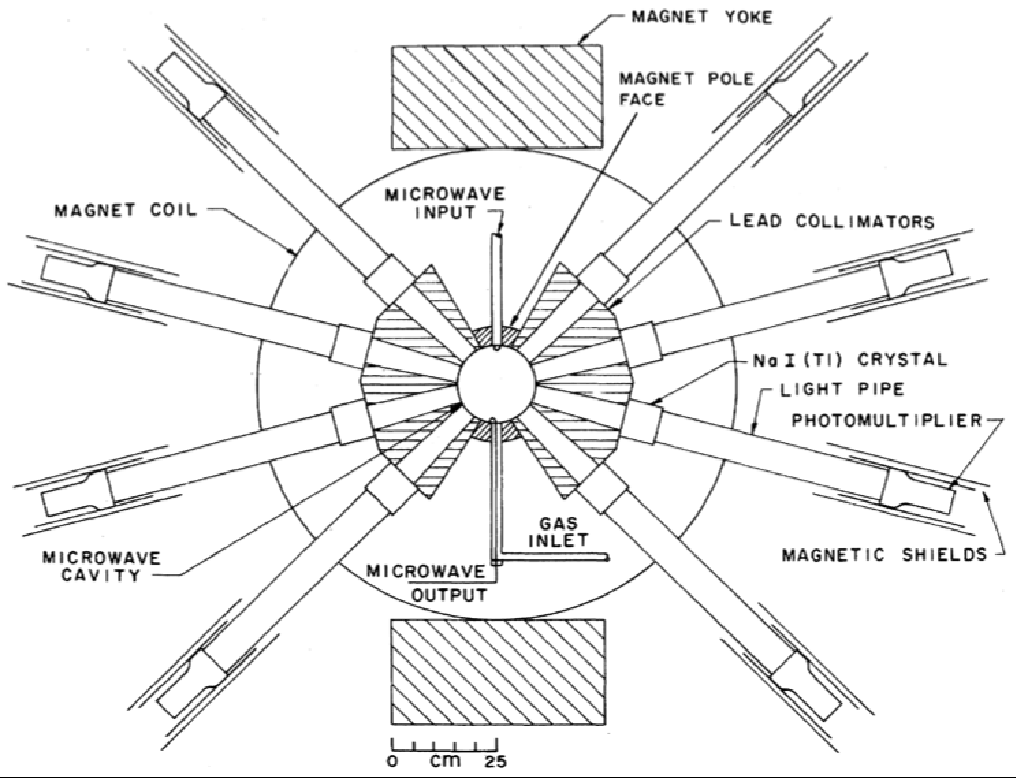}
\end{center}
  \caption{(a)The energy level of Ps as function of the static magnetic field. 
$\Delta \nu$ represents the HFS, and o-Ps state ($^{3}S_{1}$) is higher 
than p-Ps ($^{1}S_{0}$). 
Arrow shows the Zeeman shift between $M_z=0$ and $M_z=\pm1$ at B=8kgauss.
(b) The experimental setup using the Zeeman shift~\cite{HFSnew}.}
\end{figure}

The experimental setup used in Ref.~\cite{HFSnew} shows in Fig.1(b).
Positronium was produced with the $\beta^+$ source and gas($N_2$, Argon etc)
in the RF cavity, in which the high power 2.3GHz microwave was stored.
Static magnetic field (about 8~kgauss) was also applied
and scanned near the resonance of the Zeeman shift.
When the energy of the Zeeman shift is just on the RF frequency, 
the transition from $m_z=\pm$1 to $m_z$=0 increases, and the $m_z$=0 state decays into
2$\gamma$ immediately through p-Ps state.
The 2$\gamma$ decay, which were tagged with back-to-back topology using NaI scintillator shown in Fig.1(b), 
increases on the resonance. 
HFS can be determined as center value of the resonance peak 
by scanning the magnetic field.
Since Ps is produced in the gas, the Ps collides with the gas molecule and
the electric field of the gas molecule makes the shift of the energy 
state called as the Stark effect, 
and about 10ppm shift was observed for 1 atm gas\cite{HFSnew}. 
The HFS in gas were measured by changing the gas pressures. 
The measured values were extrapolate to zero pressure,
and the HFS in the vacuum was obtained.

There are two possibilities of the systematic errors in this method.
\begin{enumerate}
\item The Ps is widely spread in the RF cavity (size 17cm in diameter) 
and the size of the used magnet were limited comparing the cavity as shown in Fig.1(b).
Non-uniformity of the magnetic field filed was worse than about 10ppm level and
the correction has been applied in the analysis.
But the errors of the magnetic field is enhanced and propagated on the final HFS results
by factor 2. 
So Non-uniformity of the magnetic field might be the unknown systematic errors.
\item 2nd is the effect of the non-thermalized o-Ps, which is
the same as in the decay rate measurement\cite{pslife}.
Extrapolation procedure is assumed that Ps is well thermalized and 
the mean velocity of the Ps is the same for the various pressured gases.
We have already shown in the decay rate measurements\cite{pslife}
that this well-thermalized assumption is not satisfied, 
and that this extrapolation made the serious systematic errors, 
known as "o-Ps lifetime puzzle". 
The non-thermal Ps would affect also on the HFS measurement.
\end{enumerate}

\section{Direct transition method and Gyrotron}
 
Directly measurement of the HFS transition without a static magnetic field is 
free from the systematic error from the magnetic field mentioned in (1).
A powerful 203~GHz radiation field is necessary to cause the direct transition,
since this transition is suppressed (transition rate is $3\times 10^{-9} s^{-1}$). 
We are developing sub-THz to THz light source called a gyrotron, and also a high-finesse
Fabry-P\'erot cavity to accumulate sub-THz photons for the direct HFS measurement.
Developing the high quality/high power source for the  (sub) THz  region
is useful and the interesting for both science and technology.

The gyrotron\cite{Idehara} is a novel high power radiation source for sub THz to THz frequency region.
The electrons are produced and accelerated at the DC electron gun, concentrated and rotated as cyclotron
motion in the superconducting magnet. The cyclotron frequency $f_c$ is
\begin{equation}
	f_c = \frac{eB}{2\pi{}m_0\gamma},
\end{equation}
where $B$ is the magnetic field strength, $m_0$ is the electron rest mass and
$\gamma$ is the relativistic factor of the electron.
A cavity is placed at the maximum magnetic field, whose resonance frequency is tuned just
to the cyclotron frequency.
The electrons stimulate resonance of the cavity and produce coherent photons at the cavity.
The photons are guided to the output port through the window, while electrons are dumped
at the collector.
We developed a gyrotron operating at $f_c = 203$ GHz with $B=7.425$ Tesla, $\gamma \sim 1.02$.
Figure~2(a) shows the photograph of the gyrotron under developing for the HFS measurement.(Gyrotron FU CW V). 

\begin{figure}[h]
\begin{center}
\includegraphics[height=60mm]{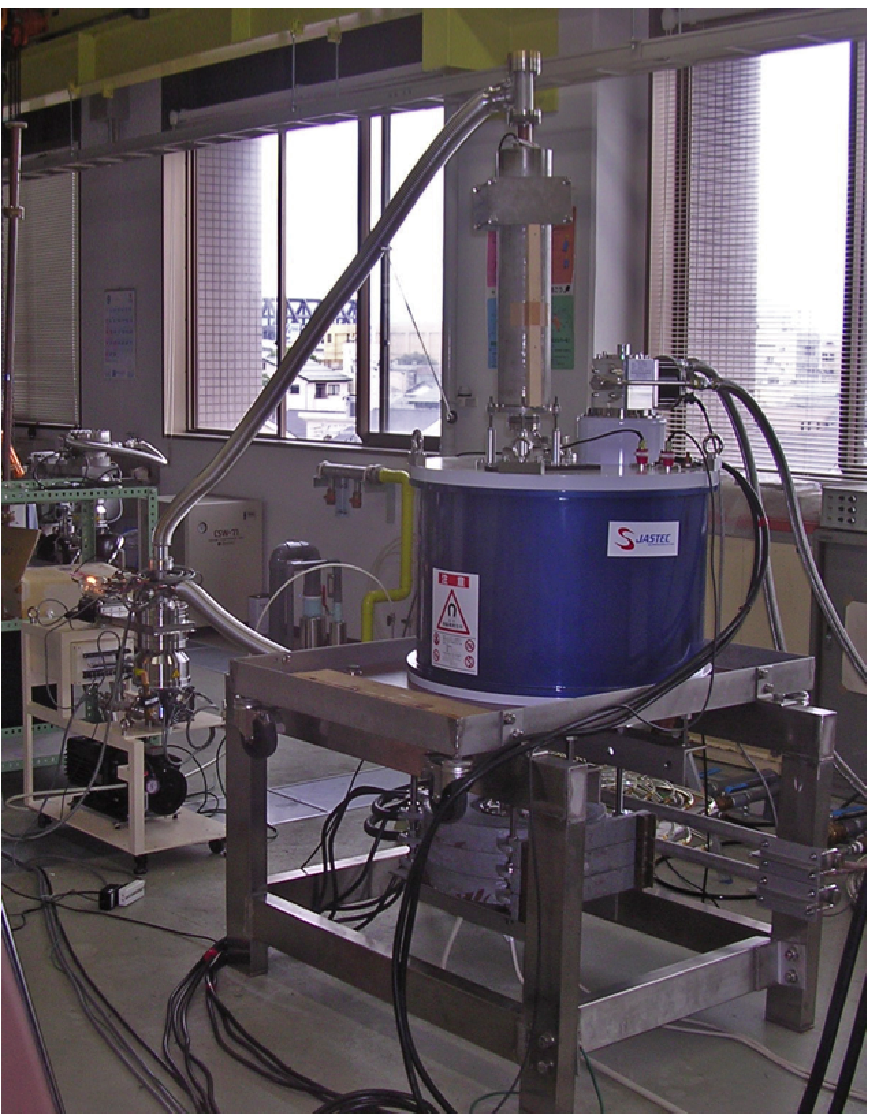}
\includegraphics[height=60mm]{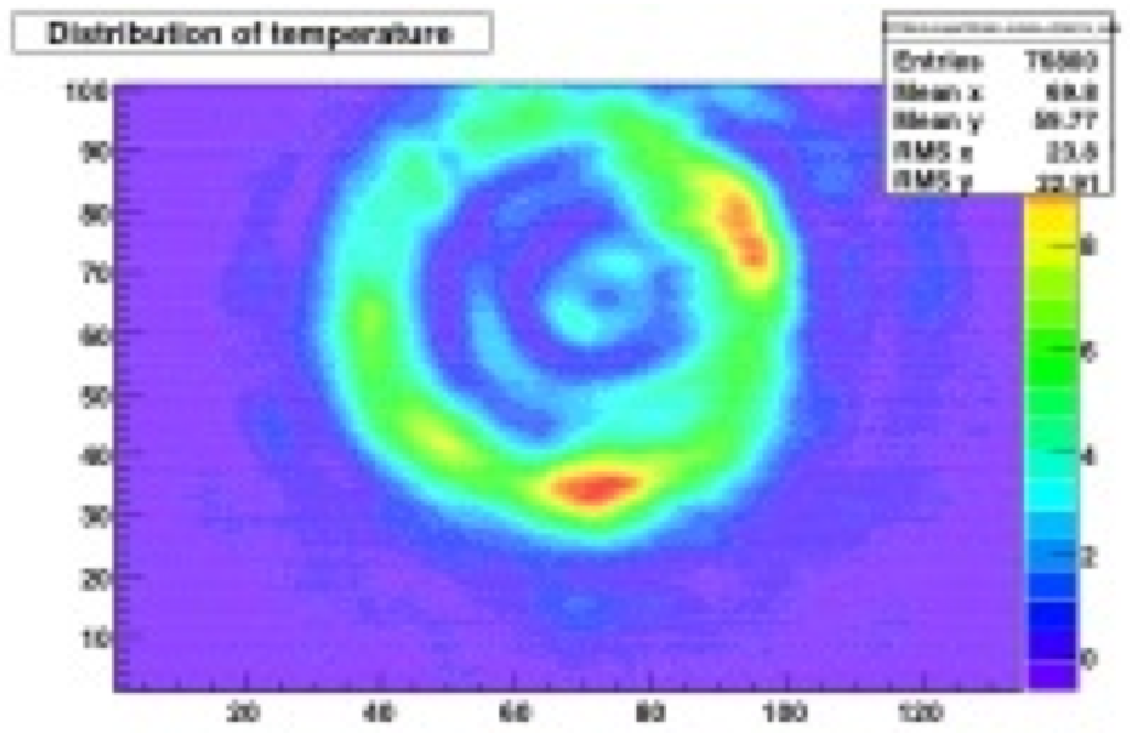}
\end{center}
  \caption{(a) Photograph of the Gyrotron FU CW V (b) The power shape measured with the infrared camerae
at the  output of the waveguide.}
\end{figure}

The performance requested to the light source are listed here:
\begin{enumerate}
\item The power is high $> O(100)$ W.
The obtained radiation power of the gyrotron is 609 W at the window, 
which is reduced to 440 W during transmission through the waveguide system to the positronium cavity.
New effective transfer system using Gaussian converter is under developing~\cite{suehara}. 
\item The powerful light can be produced continuously (CW mode)
and the stable operation is need for the long time.
The power of light should be controlled and monitored with an accuracy of $O(0.1)$\%.
The improvement stability of the power is in progress using the feedback system to initial electron-gun.
\item The tunable frequency is also big challenge of (sub) THz light source. 
The frequency can be tuned by changing the $\gamma$ factor with different acceleration of electrons,
but the tuning range is limited by the resonant width of the cavity to several hundreds of MHz.
Another mechanism to make the frequency tune is developed using the Backward-wave Oscillator~\cite{BWO}.
\item The stability on the frequency is also necessary with $O(1)ppm$-level.
The spectral width is determined by $B$ uniformity and $\gamma$ spread by thermal distribution of
electrons, and is expected to be less than MHz, which is narrow enough to make resonance
at the Fabry-P\'erot cavity. Measurement results of a similar gyrotron shows
the spectral width is less than 10 kHz.
\end{enumerate}

The power distribution measured with the infrared camera is shown in Fig.2(b), and
the wave form is TE03 mode, which resonance mode is used in the gyrotron.
This TE03 mode is changed into the Gaussian mode (1) to transfer power without loss
(2) to couple with the Fabry-P\'erot cavity effectively.
We use Gaussian-Converter using Vlasov mirror with two other mirrors.
The detail is summarized in note~\cite{suehara}.

\section{Experimental setup}

Figure 3(a) shows the schematic view of the experimental setup.
$^{22}Na$ positron source(1MBq) is installed in the 
thin plastic scintillator(100$\mu$m),
and the timing information of the positron emission
is tagged (positron emits at t=0).
The cavity is filled with mixed air of nitrogen or iso-C$_4$H$_{10}$ 
to form positronium atoms (20\% efficiency).
p-Ps (25\% of all Ps) annihilates to two 511 keV gammas immediately 
as well as positron annihilation,
while o-Ps (75\%) remains with $\tau \sim 142$ ns 
and decays to three photons ($<$ 511 keV), generating delayed signal at LaBr$_3$
scintillator.

Six LaBr$_3$ scintillators surround the cavity to catch photons 
with energy resolution of $\sim 4$\% which can efficiently
separate 511 keV photons (evidence of HFS transition) from photons from o-Ps decay.
The LaBr$_3$ scintillators have timing resolution of $\sim 300$ psec to separate delayed events (signal)
from prompt events (positron annihilation).

There are two promising benefits to use time information 
and this is 2nd wonderful technique of our new method.
\begin{enumerate}
\item Many part of positron(~80\%) is just annihilated into 2$\gamma$ at t=0,
and this 2$\gamma$ annihilation events make S/N seriously worse.
The 2$\gamma$ annihilation background events can be removed 
dramatically with the requirement of t$>$10nsec.
\item Thermalization process can be determined with the energy spectrum 
measured as the same as in the measurement of the decay rate\cite{pslife}.
the non-thermalization effect are measured directly.
2nd systematic error can be removed.
\end{enumerate}

\begin{figure}[h]
\begin{center}
\includegraphics[height=45mm]{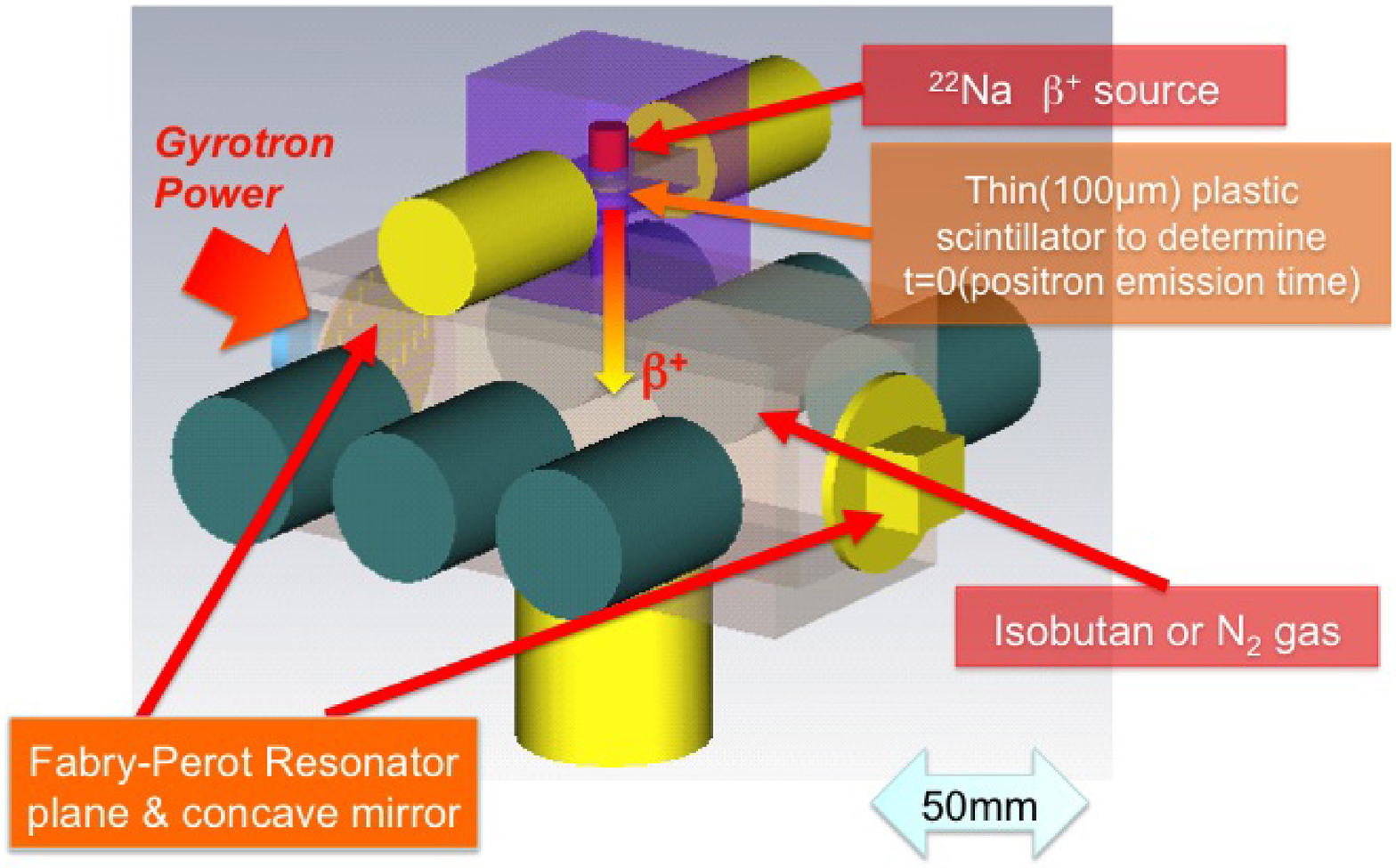}
\includegraphics[height=45mm]{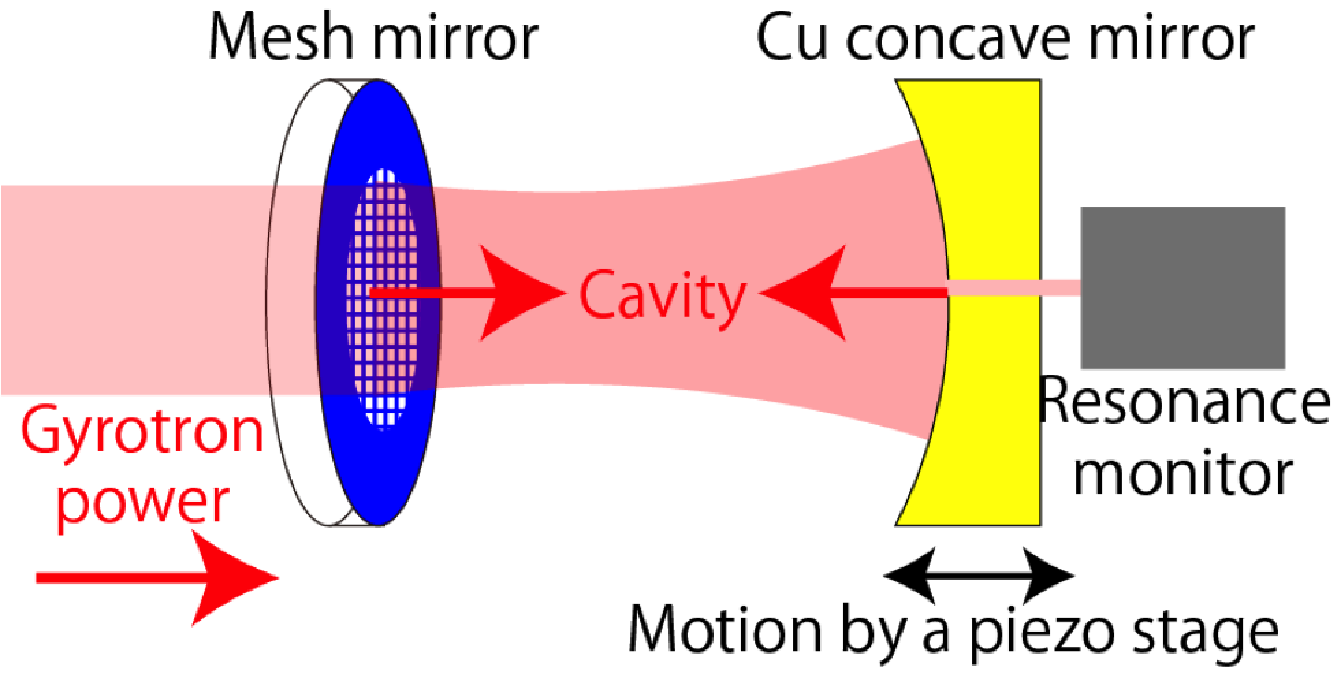}
\end{center}
  \caption{(a) The schematic view of the experimental setup (b) Fabry-Perot Cavity}
\end{figure}

\section{Fabry-P\'erot cavity}

Photons produced at the gyrotron are transported and accumulated in a cavity to
cause the Ps-HFS transition. 
Since 203 GHz ($\lambda = 1.475$ mm) photons can be treated optically
at the centimeter (or larger) size scale, we plan to use a Fabry-P\'erot cavity as shown 
in Fig.3(b).
It consists of two opposing mirrors to confine photons between them.
We use a metal-mesh mirror(Fig.4(a)) on the input side of the cavity and
a copper concave mirror(Fig.4(b)) on the other side.
A concave mirror is mounted on a piezo stage to shift cavity-length precisely (step 200$nm$).  
Photon power is stored when the cavity in the resonance for the input sub-THz wave.
Input, transmitted and reflected powers were monitored by three pyroelectric power monitors.

\begin{figure}[h]
\begin{center}
\includegraphics[width=60mm]{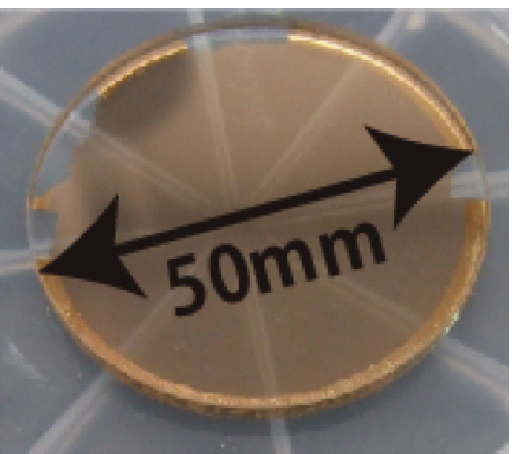}
\includegraphics[width=60mm]{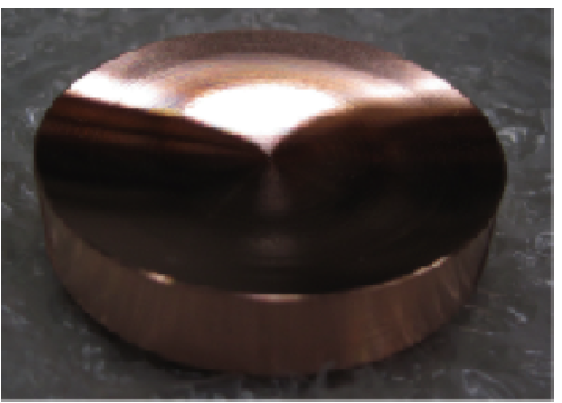}
\end{center}
  \caption{Photograph of the prototype mirrors (a) plane mirror(SiO$_2$) with the mesh(Au) (b) concave mirror(Cu)}
\end{figure}

The two most important characteristics of cavities are
``finesse" and ``input coupling".
Finesse can be written as $\mathcal{F} = 2\pi/(1-\rho)$,
where $\rho$ is the fraction of power left after one round-trip,
characterizing the capability of the cavity to store photons.
There are three sources of the power losses; 
(1) diffraction loss, (2) medium loss and (3) ohmic loss.
The first two losses are expected to be negligible in our cavity.
The last one, ohmic loss, occurs at the cavity mirrors, 
which is around 0.15\% at the copper mirror and more at the mesh mirror.
The ohmic loss at the mesh mirror varies by mesh parameters,
which can be calculated by field simulation (see Table.1).
The ohmic loss and the coupling performance are opposed to each other and 
we have to optimize the best parameters of the mesh.
The finesse and the input coupling were measured (Fig.5(a)) using various mesh and 
concave mirror parameters to check the calculated results,
and the Breit-Wigner resonance shape is observed as shown in Fig.5(b).
Finesse could be obtained from the width of the resonance 
with the transmit power monitor, and the input coupling was seen by the reflection monitor.
With current best combination of the mesh and the concave mirror, $\mathcal{F} = 650$ was obtained.
For the input coupling, concrete value could not be 
obtained because of interference of the reflection power
and difficulty to determine absolute power of the reflection/transmission 
because of non-optimal setup of the power measurements.

\begin{table}[h]
\caption{\label{meshparam}
Mesh parameters with simulation and measurement result.
reflectance and transmittance are from estimations by the simulation,
and finesse is from results of the measurements.
}
\begin{center}
\begin{tabular}{rrrrrr}
mesh material & line width & line separation & reflectance & transmittance & finesse\\ \hline
gold   &  20 $\mu$m &  50 $\mu$m & 99.3\% & 0.32\% & 650\\
gold   &  10 $\mu$m &  50 $\mu$m & 98.6\% & 0.75\% & 290\\
silver &  50 $\mu$m & 130 $\mu$m & 96.9\% & 2.70\% & 180\\ \hline
\end{tabular}
\end{center}
\end{table}

\begin{figure}[h]
\begin{center}
\includegraphics[width=60mm]{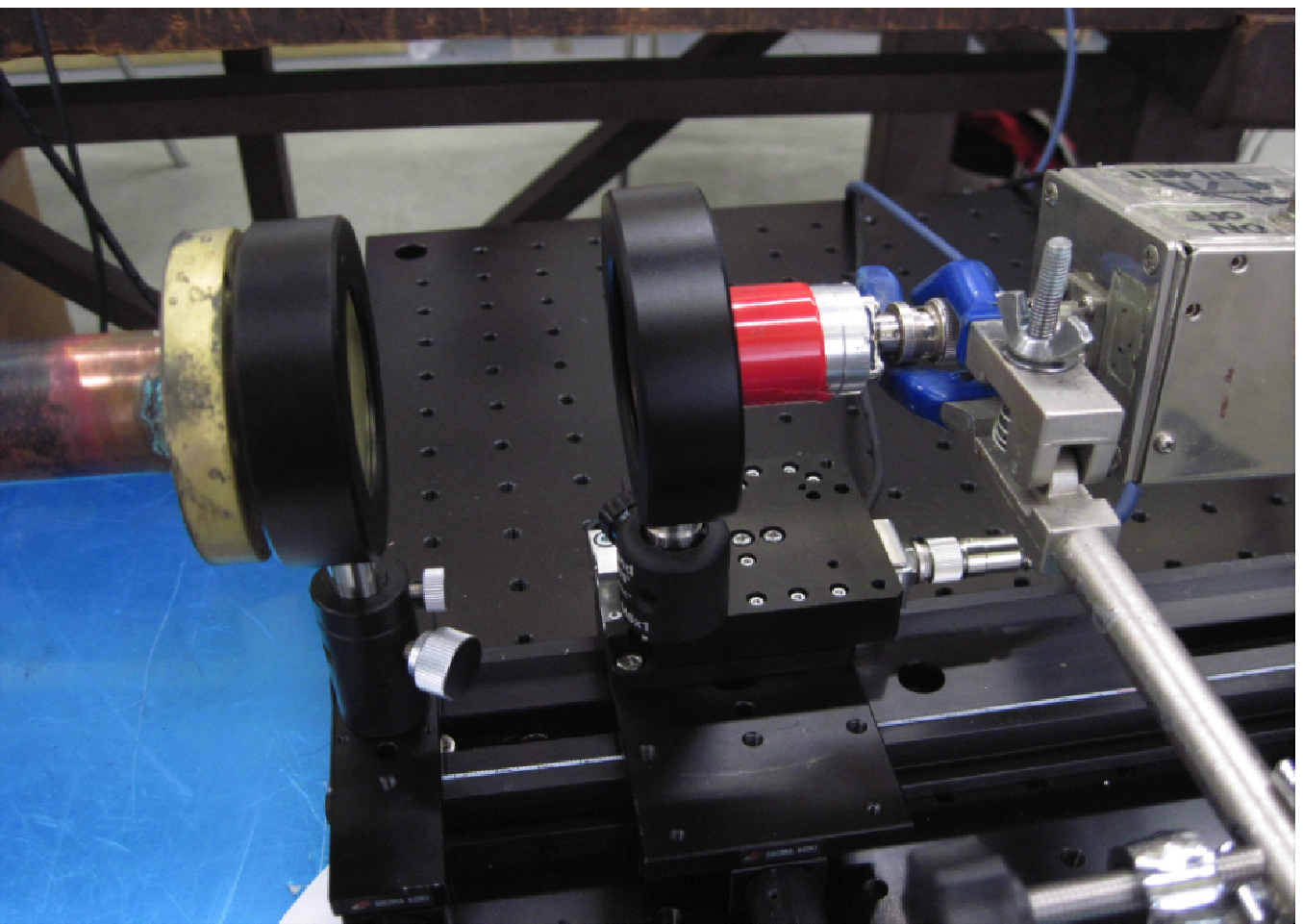}
\includegraphics[width=60mm]{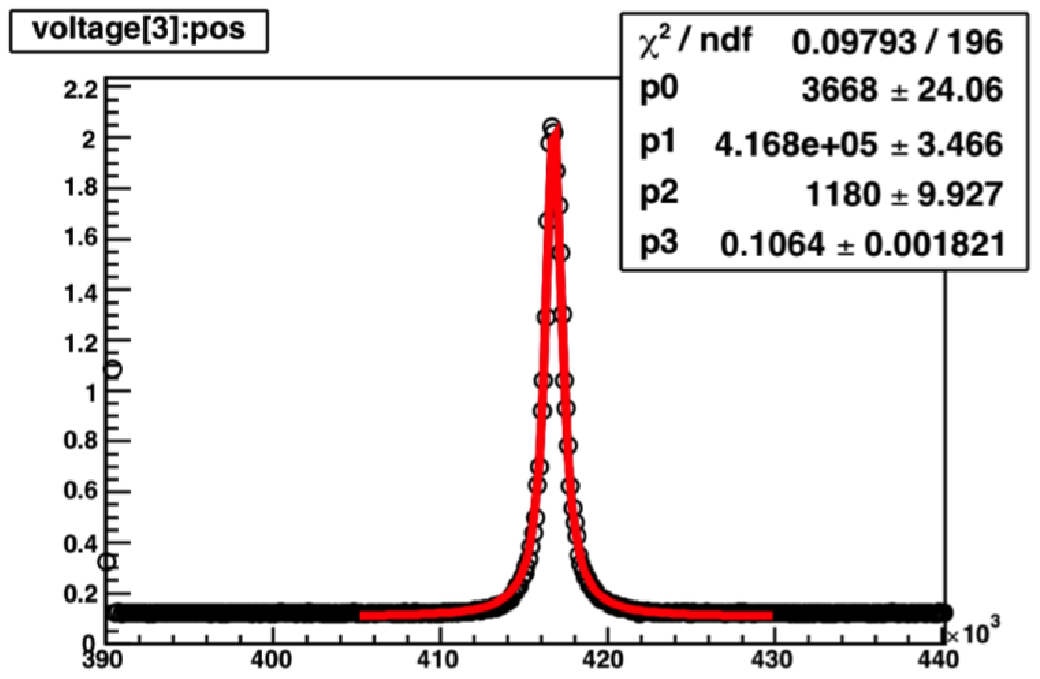}
\end{center}
  \caption{(a) The photograph of prototype of Pabry-P\'erot cavity
           (b) The BW resonance measured with (a)}
\end{figure}

\section{Summary}

There is large discrepancy of 3.9$\sigma$ between  the measured HFS values 
and the QED prediction ($O(\alpha^3)$).
We point out the possible unknown systematic errors in the previous experiments.
and we propose new method to measure HFS precisely and directly.
A gyrotron, a novel sub-THz light source is used with a high-finesse 
Fabry-P\'erot cavity to obtain enough radiation power at 203 GHz.
The present status of the optimization studies and current design of the
experiment are summarized in this note.

\section*{Acknowledgement}



\begin{thebibliography}{9}
\bibitem{HFSnew} A.P.Mills, Phys. Rev. A \textbf{27} 262 (1983), and 
M.W.Ritter et.al., Phys. Rev. A \textbf{30} 1331 (1984).
\bibitem{HFSth} K.Melnikov et.al., Phys. Rev. Lett. \textbf{86} 1498 (2001), 
B. A.Kniehl and A.A.Penin  Phys. Rev. Lett. \textbf{85} 5094 (2000),
and R.J.Hill,  Phys. Rev. Lett. \textbf{86} 3280 (2001).
\bibitem{pslife} S.Asai et al., Phys. Lett. B. \textbf{357} 475 (1995).
\bibitem{Idehara} T.Idehara et al.,IEEE Trans. Plasma Sci. \textbf{27} 340 (1999).
\bibitem{suehara} T.Suehara shows in the same workshop. 
\bibitem{BWO} T.H.Chang and T.Idehara et al.,J.Appl. Phys. \textbf{105} 063304 (2009).

\end{thebibliography}
\end{document}